\documentclass[12pt]{iopart}
\newcommand{\nnbe}{\stackrel{{\scriptscriptstyle (} - {\scriptscriptstyle )}}{\nu}\!\!\!\! _{\rm e}}
\usepackage{iopams}
\usepackage{epsfig}
\begin{document}

\title[Probing Low Energy Neutrino Backgrounds]{Probing Low Energy Neutrino Backgrounds with Neutrino Capture on Beta Decaying Nuclei}

\author{Alfredo G Cocco$^1$, Gianpiero Mangano$^1$ and Marcello Messina$^2$}

\address{$^1$ Istituto Nazionale di Fisica Nucleare - Sezione di
Napoli - \\ Complesso Universitario di Monte S.Angelo, I-80126
Napoli, Italy
\\
$^2$ Laboratorium f\"ur Hochenergiephysik - Universit\"at Bern\\
Sidlerstrasse 5, CH-3012 Bern, Switzerland}

\ead{alfredo.cocco@na.infn.it\\ mangano@na.infn.it \\
marcello.messina@cern.ch}

\begin{abstract}
We study the interaction of low energy neutrinos on nuclei that
spontaneously undergo beta decay showing that the product of the
cross section times neutrino velocity takes values as high as
$10^{-42}$ cm$^2$ $c$ for some specific nuclei that decay via
allowed transitions. The absence of energy threshold and the value
of the cross section single out these processes as a promising
though very demanding approach for future experiments aimed at a
direct detection of low energy neutrino backgrounds such as the
cosmological relic neutrinos.
\end{abstract}


\maketitle

\section{Introduction}

The interaction of an electron (anti)neutrino with a nucleus $N$
naturally undergoing beta (positron) decay to the daughter nucleus
$N'$
\begin{equation}
\nnbe + N \rightarrow N' + e^\pm \label{main} \, ,
\end{equation}
shows the remarkable property of having no energy threshold on the
value of the incoming neutrino energy. Indeed, in these cases the
energy balance of the corresponding beta decay reactions is such
that $M(N)-M(N')=Q_\beta>0$ with $M(N)$, $M(N')$ the mass of
neutral atoms. Neutrino interaction of this type is thus always
energetically allowed, no matter the value of the incoming
neutrino energy $E_\nu$. Reaction~ (\ref{main}) will be denoted in
the following as Neutrino Capture on Beta decaying nuclei (NCB).
In the limit of vanishing values of neutrino mass $m_\nu$ and
$E_\nu$ the neutrino contributes to (\ref{main}) uniquely via its
lepton flavor quantum number and in this case the electron in the
final state has exactly the beta decay endpoint energy $Q_\beta$.
However, for finite $m_\nu$ the electron kinetic energy is
$Q_\beta+E_\nu \geq Q_\beta + m_\nu$, while electrons emerging
from the analogous beta decay has at most an energy $Q_\beta -
m_\nu$, neglecting nucleus recoil energy. A minimum gap of
2$m_\nu$ is thus present and this at least in principle allows to
distinguish between beta decay and NCB interaction.

In neutrino physics, NCB represents to date the only known
reaction able to unambiguously detect electron (anti)neutrino
having arbitrary low energies. Neutrinos with energies between 10
and 100 keV are still undetected even if some measurements have
been proposed~\cite{trofi, gioma, mclau, lens}. Low energy
neutrino physics is indeed crucial to clarify our picture of
fundamental particle interactions. Compelling results about
neutrino mixing~\cite{pdg06} and the corresponding non zero
neutrino mass evidence raise a series of questions that could be
addressed with the aid of the NCB process. The value of the
absolute neutrino mass scale and the existence of low mass sterile
neutrino mixing with ordinary flavors are examples of what can be
done exploring this low energy regime.

The idea of using NCB to measure the cosmological relic neutrino
background predicted in the framework of the Hot Big Bang model
was already advocated many years ago in \cite{weinberg}. The
original idea was that if relic neutrinos have a large chemical
potential $\mu$, then both beta decays and NCB would show a
depletion in the electron (positron) energy spectrum in a gap of
order $\mu$ around the zero neutrino mass endpoint $Q_\beta$.
Presently, we know that these neutrinos have a number density of
order $n_\nu \sim 50$ cm$^{-3}$ neutrino (or antineutrino) per
flavor and are characterized by a very small mean kinetic energy,
of the order of 6.5 $T_\nu^2/m_\nu$ or 3.15 $T_\nu$ for
nonrelativistic and relativistic neutrinos, respectively with $
T_\nu = (4/11)^{1/3} T_\gamma  \sim 1.7 \cdot 10^{-4}$ eV. Big
Bang Nucleosynthesis constrains relic neutrino--antineutrino
asymmetry, i.e. the ratio $\mu/T_\nu$ for electron neutrino to be
very small $\mu/T_\nu \leq 0.1$, see e.g. \cite{cuoco,pat}, and
this bound also applies to all active neutrino flavors because of
the oscillation mechanism \cite{dolgov}. This implies that, unless
more exotic scenarios are considered, as a larger amount of
relativistic degrees of freedom in the Early Universe, the effect
of neutrino degeneracy in beta decays and NCB is too small to be
detected experimentally. However, as we mentioned already for
massive neutrinos a similar gap around $Q_\beta$ is expected of
the order of twice the neutrino mass, which for $m_\nu \sim 1$ eV
is several orders of magnitude larger than the corresponding
effect due to neutrino-antineutrino asymmetry.

Presently, neutrino mass in the eV range is still allowed by data.
Indeed, oscillation experiments only provide a lower limit to the
mass of (at least) one neutrino mass eigenstate of the order of
0.05 eV \cite{valle}, while direct measurements of electron energy
spectrum in $^3$H decay gives $m_\nu< 2$ eV
\cite{tritium1,tritium2}. A large improvement in this respect will
be provided by the KATRIN experiment, whose expected sensitivity
is 0.2 eV \cite{katrin}. On the other hand, cosmological data from
Cosmic Microwave Background anisotropies and Large Scale Structure
power spectrum provide an independent bound on the sum of neutrino
masses which depending on the particular model adopted and the
number of free parameter lies in the range $0.3 \div 2 $ eV, see
e.g. \cite{lesg} for a recent review. As we will argue in the
following, if $m_\nu$ is in the eV range, future NCB experiments
could represent an almost unique way to detect cosmological
neutrinos.

It is also worth mentioning that there are several possible
sources of low energy neutrino fluxes which might be fruitfully
studied using NCB. One example is provided by the thermal neutrino
flux from the Sun, due to Compton pair production, free-bound
electron transition and plasmon decay, which produce a
neutrino/antineutrino flux in the energy range 1 eV - 10 KeV, with
an integrated flux of the order of $10^6 - 10^7$ cm$^{-2}$
s$^{-1}$ \cite{haxton}. Neutrino diffused fluxes in the 10 KeV -
1MeV energy range are instead expected from galactic and
extra-galactic star nuclear burning \cite{fiorentini}, as well as
from POP III stars \cite{iocco} and for this latter case, their
detection may represent a first direct probe of this yet unseen
metal-free first generation of stars.

Of course, the strength of neutrino interaction with ordinary
matter is very small and NCB makes no exception. In order to
design an experiment probing yet unexplored neutrino backgrounds a
detailed estimate of the cross section is mandatory. We consider
in the following a method to evaluate NCB cross section that makes
use of beta decay observables in order to minimize the
uncertainties due to nuclear matrix element evaluation. This
procedure allows to obtain an accurate numerical estimate of the
cross section in a large variety of cases. When possible, a
comparison with previous calculations made for some specific
nuclei by previous authors~\cite{holz,wale,sero} will also be
performed, though NCB interaction in the low energy neutrino limit
has not yet been systematically studied in literature.

This paper is organized as follows. In Section 2 we recall beta
decay formalism and derive the expression for the NCB cross
section which is then evaluated for several cases in Section 3. In
Section 4 the NCB to beta decay rate ratio is studied for the
particular case of the cosmological relic neutrino background.
Finally, in Section 5 we draw our concluding remarks.

\section{Neutrino cross section on $\beta^\pm$ decaying nuclei}

NCB and its corresponding beta decay are essentially the same
phenomenon. In the case of unobserved polarization the two
processes have the same invariant squared amplitude. In order to
derive an expression for the NCB cross section we make use of the
beta decay formalism, evaluating invariant amplitudes using the
description of the fermions in terms of spherical waves and the
computation of the corresponding nuclear state transition matrix
elements. This approach is based on the so called long wavelength
limit approximation $p_\nu R \ll 1$, where $p_\nu$ is the momentum
of the incoming neutrino and $R$ the nuclear radius. Due to the
typical values of $R$, this approximation holds for NCB if the
energy of the incoming neutrino satisfies the relation $E_\nu
\lesssim 10$ MeV.

Assuming an isotropic neutrino flux, the NCB integrated rate is
given by (hereafter we use natural units)
\begin{equation}
\lambda_\nu = \int \sigma_{\scriptscriptstyle {\rm NCB}} v_\nu
\,f(p_\nu) \, \frac{d^3 p_\nu}{(2 \pi)^3} \, , \label{sigma1}
\end{equation}
where $\sigma_{\scriptscriptstyle {\rm NCB}}$ is the NCB cross
section, $v_\nu$ and $p_\nu$ the modulus of neutrino velocity and
momentum, respectively and $f(p_\nu)$ the distribution function.
For example, in the particular case of cosmological relic
neutrinos it keeps a standard Fermi-Dirac form, $f(p_\nu)=
\left[\exp(p_\nu/T_\nu)+1 \right]^{-1}$ up to very small non
thermal corrections \cite{pastor}. Following the formalism
of~\cite{behrens}, for spin averaged initial state and unobserved
polarization we have
\begin{equation}
\sigma_{\scriptscriptstyle {\rm NCB}} v_\nu = \frac{G_\beta^2}{
\pi} p_e E_e F(Z,E_e) C(E_e,p_\nu)_\nu \, , \label{sigma2}
\end{equation}
where $F(Z,E_e)$ is the Fermi function. With $E_e$ and $p_e$ we
denote the energy and momentum of the outgoing electron
respectively, where
\begin{equation}
E_e = E_\nu + Q_\beta + m_e = E_\nu + m_\nu + W_o \, ,
\label{relation}
\end{equation}
with $W_o$ the corresponding beta decay endpoint. Finally, the
rate can be expressed as an integral over the electron energy
\begin{eqnarray}
\lambda_\nu &=& \frac{G_\beta^2}{2 \pi^3} \int_{W_o+2
m_\nu}^\infty p_e E_e F(Z,E_e) C(E_e,p_\nu)_\nu \nonumber \\ & &
\cdot E_\nu p_\nu \,f(p_\nu) \, d E_e \, , \label{sigma1bis}
\end{eqnarray}
where Eq. (\ref{relation}) should be used to express neutrino
energy and momentum as function of $E_e$.

The NCB rate contains the nuclear shape factor $C(E_e,p_\nu)_\nu$,
an angular momentum weighted average of nuclear state transition
amplitudes, which depends upon the nuclear properties of the
parent and daughter nuclei, thus involving the calculation of
nuclear matrix elements. Its general expression is,
see~\cite{behrens} for further details
\begin{eqnarray}
C(E_e,p_\nu)_\beta &=& \sum_{k_e,k_\nu,K} \lambda_{k_e} \Big[
M_K^2(k_e, k_\nu)
+ m_K^2(k_e, k_\nu)  \nonumber \\
&-& \frac{2\mu_{k_e} m_e \gamma_{k_e}}{k_e E_e}M_K^2(k_e,
k_\nu)m_K^2(k_e, k_\nu) \Big] \, , \label{shapef}
\end{eqnarray}
where $k_e$ and $k_\nu$ are the electron and neutrino radial wave
function indexes ($k=j+1/2$), $K$ represents the nuclear
transition multipolarity ($\left| k_e - k_\nu \right| \leqslant K
\leqslant k_e + k_\nu$) and $M_K^2(k_e, k_\nu)$, $m_K^2(k_e,
k_\nu)$ are nuclear form factor functions, which are typically
evaluated using N-body nuclear transition matrix elements in the
impulse approximation. This represents the main source of
uncertainty for $\sigma_{\scriptscriptstyle {\rm NCB}}$.

On the other hand, NCB rate is strongly related to the
corresponding beta decay process, whose rate is given by
\begin{equation}
\lambda_\beta =   \frac{G_\beta^2}{2 \pi^3} \int_{m_e}^{W_o} p_e
E_e F(Z,E_e) C(E_e,p_\nu)_\beta E_\nu p_\nu \, d E_e \, ,
\label{ratedecay}
\end{equation}
a simple relation holding  between the beta decay and the NCB
shape factors~\cite{behrens}
\begin{equation}
C(E_e,p_\nu)_\nu = C(E_e,-p_\nu)_\beta \, , \label{cpos}
\end{equation} though both variables have different kinematical
domains in the two processes.

The beta decay rate provides a relation expressing the mean shape
factor, defined as
\begin{equation}
\overline{C}_\beta = \frac{1}{f} \int_{m_e}^{W_o} p_e E_e F(Z,E_e)
C(E_e,p_\nu)_\beta E_\nu p_\nu  dE_e \, ,
\end{equation}
in terms of observable quantities, $W_o$ and the half-life
$t_{1/2}$
\begin{equation}
 ft_{1/2} = \frac{2\pi^3 \ln2}{G_\beta^2 \ \overline{C}_\beta} \, ,
 \label{ft}
\end{equation}
where $f$ is defined as
\begin{equation}
f = \int_{m_e}^{W_o} F(Z,E_e) p_e E_e  E_\nu p_\nu dE_e \, .
\end{equation}
We therefore obtain
\begin{equation}
\sigma_{\scriptscriptstyle {\rm NCB}} v_\nu = 2\pi^2 \ln2\  p_e
E_e F(Z,E_e) \frac{C(E_e,p_\nu)_\nu}{ft_{1/2} \
\overline{C}_\beta} \, , \label{sigma2bis}
\end{equation}
where the two factors $C(E_e,p_\nu)_\nu$ and $\overline{C}_\beta$
depend upon the same nuclear transition matrix elements.

It is useful to define the quantity ${\cal A}$ as
\begin{equation}
{\cal A} = \frac{f\  \overline{C}_\beta}{p_e E_e F(Z,E_e)
C(E_e,p_\nu)_\nu} \, ,
\end{equation}
and note that it contains the ratio of NCB and beta decay shape
factors
\begin{equation}
{\cal A} = \int_{m_e}^{W_o}
\frac{C(E'_e,p'_\nu)_\beta}{C(E_e,p_\nu)_\nu} \frac{p'_e}{p_e}
\frac{E'_e}{E_e} \frac{F(E'_e, Z)}{F(E_e, Z)} E'_\nu p'_\nu dE'_e
\, , \label{aint}
\end{equation}
where a prime denotes all variables depending on $E'_e$ which
should be integrated over. The ${\cal A}$ is a function of $E_\nu$
only, once the target nucleus characterized by $Q_\beta$ and $Z$
is given. The NCB cross section times neutrino velocity can then
be conveniently written as
\begin{equation}
\sigma_{\scriptscriptstyle {\rm NCB}} v_\nu= \frac{2\pi^2 \ln2
}{{\cal A}\cdot  t_{1/2}} \, , \label{sigma3}
\end{equation}
with the product $ft_{1/2}$ characterizing the various beta decay
processes. Values of $\log(ft_{1/2})$ are reported e.g.
in~\cite{singh}.

As we will see, in some relevant cases the evaluation of ${\cal
A}$ is particularly simple so that Eq. (\ref{sigma3}) can be
computed in an exact way. In all cases where this is not possible,
systematic uncertainties affecting the nuclear matrix element
evaluation are expected to largely cancel in the shape factor
ratio appearing in ${\cal A}$, and this suggests that the use of
expression~(\ref{sigma3}) might be very useful to get a reliable
estimate of the NCB cross section.

\subsection{Superallowed transitions}

Beta decays in which the nucleus undergo a transition taking place
between members of the same isospin multiplet are called
superallowed. Due to the large superposition between initial and
final nuclear states these decays have the lowest known $ft_{1/2}$
values. In particular, if the transition is of the $0^+
\rightarrow 0^+$ type only the vector current contributes to the
decay. In the limit of CVC hypothesis the vector form factor is
given by the Fermi matrix element and the shape factor can be
written as
\begin{equation}
C(E_e,p_\nu) = \left| ^{\scriptscriptstyle V}F^{\scriptscriptstyle
(0)}_{\scriptscriptstyle 000} \right|^2 = \langle {\bf F}
\rangle^2 = (T-T_3)(T+T_3+1) \, ,
\end{equation}
where $T$ and $T_3$ are the isospin quantum numbers.

In all other transitions happening inside the same isospin
multiplet ($J^\pi \rightarrow J^\pi, J \ne 0$) both the vector and
axial form factors contribute and in this case the shape factor
can be written as
\begin{equation}
C(E_e,p_\nu) = \left| ^{\scriptscriptstyle V}F^{\scriptscriptstyle
(0)}_{\scriptscriptstyle 000} \right|^2 + \left|
^{\scriptscriptstyle A}F^{\scriptscriptstyle
(0)}_{\scriptscriptstyle 101} \right|^2 = \langle {\bf F}
\rangle^2 + (g_{\scriptscriptstyle A}/g_{\scriptscriptstyle V})^2
\langle {\bf GT} \rangle^2 \, , \label{allowed}
\end{equation}
where $g_{\scriptscriptstyle A,V}$ is the axial (vector) coupling
constant. As first approximation both form factors can be
evaluated using the initial and final state spin and isospin
wavefunctions. However, a precise calculation requires the
evaluation of higher order correction to nuclear matrix elements.

We notice that for all superallowed decay the form factors do not
depend on $E_e$. In this case ${\cal A}$ can be written as
\begin{equation}
{\cal A} = \frac{f}{p_e E_e F(Z,E_e)} \, , \label{super}
\end{equation}
and we get for NCB processes
\begin{equation}
\sigma_{\scriptscriptstyle {\rm NCB}} v_\nu = 2\pi^2 \ln2
\frac{p_e E_e F(Z, E_e)}{ft_{1/2}} \, , \label{xssup}
\end{equation}
which only depends on the half-life and the $Q_\beta$ of the
corresponding beta decay.

To illustrate the difference between expressions~(\ref{sigma2})
and~(\ref{sigma3}) we consider the specific case of Tritium which
decays to $^3{\rm He}$ with $Q_\beta=18.591(1)$ keV~\cite{audi}
and an half-life of $t_{1/2}=12.32(4)$ years~\cite{simpson}. We
adopt $\langle {\bf F} \rangle^2 = 0.9987$ and $\langle {\bf GT}
\rangle = \sqrt{3}\cdot (0.964\pm 0.016)$ from~\cite{schiavilla}
assuming a total 1.6\% systematic uncertainty on the Gamow-Teller
matrix element evaluation. Using expression~(\ref{sigma2}) and
(\ref{allowed}) with $G_{\rm F}=1.16637(1)\times 10^{-5}$
GeV$^{-2}$, $g_{\scriptscriptstyle A} = 1.2695(29)$ and $\left|
V_{ud} \right| = 0.97377(27)$~\footnote{We recall here that errors
on $g_{\scriptscriptstyle A}$ and $V_{ud}$ are strongly
correlated.} \cite{pdg06} we find
\begin{equation}
\sigma_{\scriptscriptstyle {\rm NCB}} (^3{\rm H})
\,\frac{v_\nu}{c} = \left( \ 7.7 \pm 0.2\  \right) \times 10^{-45}
\ {\rm cm}^2 \, ,
\end{equation}
in case of massless neutrinos and in the limit of $p_\nu
\rightarrow 0$. We notice here that the main contribution to the
quoted error is due to the uncertainty on the evaluation of the
Gamow-Teller matrix element. This result is in agreement with
previous calculations~\cite{holz, grotz}. On the other hand,
evaluating $\sigma_{\scriptscriptstyle {\rm NCB}} (^3{\rm H})
\,v_\nu$ using expression~(\ref{xssup}) gives
\begin{equation}
\sigma_{\scriptscriptstyle {\rm NCB}} (^3{\rm H}) \frac{v_\nu}{c}
= \left( \ 7.84 \pm 0.03\  \right) \times 10^{-45} \ {\rm cm}^2 \,
,
\end{equation}
where the error is entirely due to the experimental uncertainties
on $Q_\beta$ and $t_{1/2}$ and the effect of matrix element
uncertainty is now fully removed.

\subsection{Allowed transitions}

The neutrino cross section for nuclei that undergo allowed beta decay
can be evaluated only using a first order approximation since the beta decay
shape factor is written in general as
\begin{equation}
C(E_e,p_\nu)_\beta = \left| ^{\scriptscriptstyle
V}F^{\scriptscriptstyle (0)}_{\scriptscriptstyle 000} \right|^2 +
\left| ^AF^{\scriptscriptstyle (0)}_{\scriptscriptstyle 101}
\right|^2 + {\cal O}(p_e\,R){\cal O}(\alpha Z) \, .
\end{equation}
If only the leading terms are taken into account one gets the
following approximation
\begin{equation}
C(E_e,p_\nu)_\beta = C(E_e,p_\nu)_\nu = {\rm constant} \, ,
\end{equation}
and from (\ref{sigma2}) one derives an order of magnitude estimate
for the NCB interaction cross section. The effect of the higher
order terms is however, not negligible and more precise estimates
require nuclear matrix element evaluation for each specific
transition. Nevertheless, we expect that at a higher level of
accuracy with respect to the leading approximation as above can be
obtained using (\ref{sigma3}) and the fact that for the shape
factor $ratio$ is reasonable to assume
\begin{equation}
\frac{C(E_e,p_\nu)_\beta}{C(E_e,p_\nu)_\nu} \simeq 1 \, .
\label{allw}
\end{equation}
Within this approximation ${\cal A}$ is obtained using
expression~(\ref{super}).

\subsection{Unique K-th forbidden transitions}

Neutrino cross section on nuclei that decay via K-unique forbidden
beta decay takes a simple form since there is only one nuclear
form factor involved in the shape factor. In particular, its
expression can be written as
\begin{eqnarray}
&& C(E_e,p_\nu)_\beta = \left| ^{\scriptscriptstyle
A}F_{\scriptscriptstyle LL-11}^{\scriptscriptstyle (0)} \right|^2
 \nonumber \\
 && \times \sum_{n=1}^{L}  {\cal B}_L^n \lambda_n (p_e R)^{2(n-1)} (p_\nu
 R)^{2(L-n)}\, ,
\end{eqnarray}
where $K$ is the degree of forbiddeness of the decay, $L=K+1$ and
${\cal B}_L^n$ and $\lambda_n$ are respectively, numerical
coefficients and functions that can be evaluated as
in~\cite{behrens}.
In particular, we can define the following functions $u_i$ for
$i$-forbidden decays
\begin{eqnarray}
u_1(p_e, p_\nu) &=& p_\nu^2 + \lambda_2 p_e^2 \, , \\
u_2(p_e, p_\nu) &=& p_\nu^4 + \frac{10}{3} \lambda_2 p_\nu^2 p_e^2
+ \lambda_3 p_e^4 \, , \\
u_3(p_e, p_\nu) &=& p_\nu^6 + 7 \lambda_2 p_\nu^4 p_e^2 + 7
\lambda_3 p_\nu^2 p_e^4  + \lambda_4 p_e^6 \, .
\end{eqnarray}
The corresponding shape factors are then given by
\begin{equation}
C(E_e,p_\nu)_{\beta}^i = \left[ \frac{R^{i}}{(2i+1)!!} \right]^2
\left| ^{\scriptscriptstyle A}F_{\scriptscriptstyle (i+1)\, i\,
1}^{\scriptscriptstyle (0)} \right|^2 u_i(p_e, p_\nu) \, .
\end{equation}
Since there are only even powers of the neutrino momentum $p_\nu$
these expressions are valid both for beta decay and NCB
interaction. The quantity ${\cal A}$ can be written using these
expressions as
\begin{equation}
{\cal A}_i = \int_{m_e}^{W_o} \frac{u_i(p'_e, p'_\nu)p'_e E'_e
F(Z, E'_e)}{u_i(p_e, p_\nu) p_e E_e F(Z, E_e)} E'_\nu p'_\nu  {\rm
d} E'_e \, ,
\end{equation}
and results to be independent of nuclear form factors.

\section{Estimating the cross section}

We have evaluated NCB cross section for several nuclei using
parametrization of the Fermi function and of the radial wave
function coefficients $\lambda_n$ in the shape factor as in
\cite{behrens} and \cite{wilk} to account for finite nuclear size
effect. Two numerical programs have been used to evaluate
numerically $F(Z, E_e)$ and $\lambda_n$, implemented with
\verb"Mathematica"~\cite{mathematica} and C++ in a Linux
environment, using a high precision calculation of the Gamma
function~\cite{matpack}. In both cases the algorithms have been
checked against each other and compared with the values reported
in~\cite{bandb} in the range where these are available. A total
accuracy of $10^{-3}$ was achieved on $F(Z,E_e)$ and $\lambda_n$
for values of $Z$ up to $Z\sim 40$, while for higher values of $Z$
and at large values of electron (positron) energy ($E_e \gtrsim
10$ MeV) the accuracy worsen to a few percent.
\begin{figure}
\includegraphics[width=15cm]{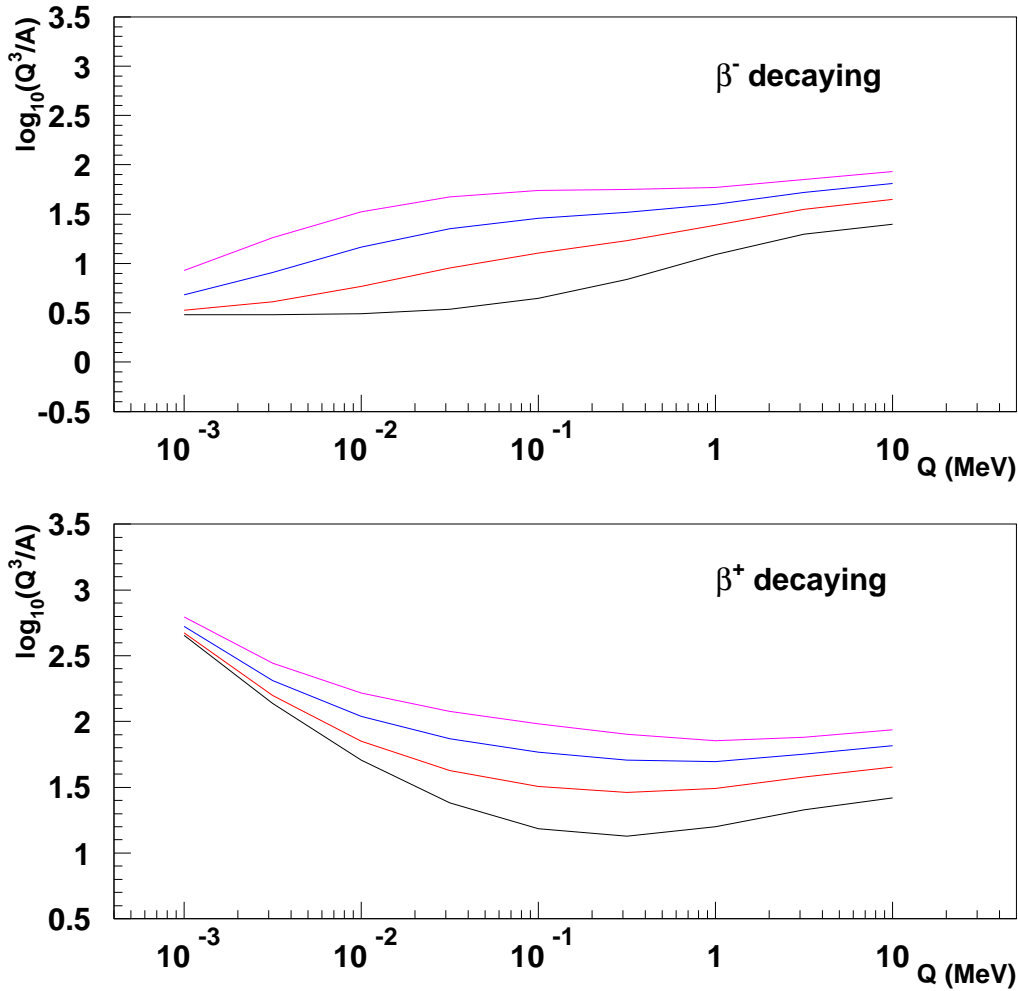}
\caption{\label{fig:aval} Values of $1/{\cal A}\cdot Q_\beta^3$
versus $Q_\beta$ evaluated for $\beta$ decaying nuclei in the
limit of $p_\nu \rightarrow 0$. The four curves represent from
bottom to top superallowed, first unique forbidden, second unique
forbidden and third unique forbidden transitions, respectively.
Curves are shown for $Z=20$ and a nuclear radius given by $R=1.2
A^{1/3}$ fm, where $A=2.5 Z$}
\end{figure}

To illustrate the behavior of $\sigma_{\scriptscriptstyle {\rm
NCB}} v_\nu$ as a function of the nucleus we first consider in
Figure~\ref{fig:aval} the ratio  $Q_\beta^3/\cal A$ for
superallowed and unique-forbidden transitions in the case of
massless neutrino and in the limit of small neutrino momentum.
Notice that as expected, this ratio is only weakly depending on
$Q_\beta$ as from (\ref{aint}) $\cal A$ grows as $Q_\beta^3$.
\begin{figure}
  \includegraphics[width=15cm]{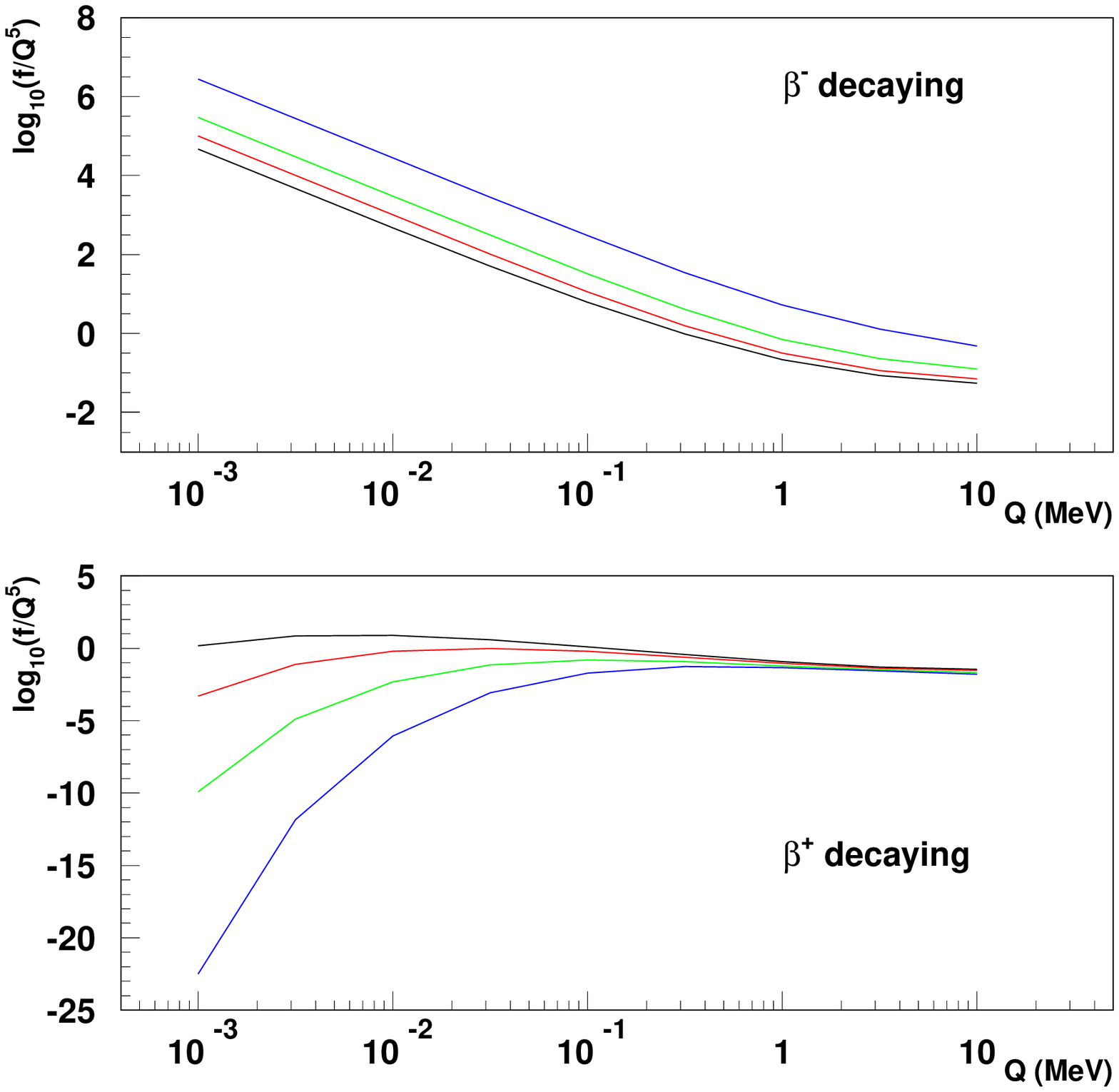}
\caption{\label{fig:fval} Values of $f\cdot 1/Q_\beta^5$ for
$\beta$ decaying nuclei in the limit of $p_\nu \rightarrow 0$. The
curves are evaluated using different values of the parent nucleus
$Z$, from bottom to top (top to bottom in case of $\beta^+$)
Z=20,40,60,80, respectively. The nuclear radius is taken to be
$R=1.2 A^{1/3}$ fm, where $A=2.5 Z$}
\end{figure}

As we mentioned in Section II, the $ft_{1/2}$ value characterizes
the particular beta decay process corresponding to the NCB. Values
of $\log(ft_{1/2})$ are reported in~\cite{singh} with typical mean
values of 3.44 for superallowed transitions, 5.5 for allowed
transitions, 9.5 for first unique forbidden transitions, 15.6 for
second unique forbidden transitions and finally, 21.1 for third
unique forbidden transitions. Using these results the NCB cross
section can be derived using (\ref{sigma3}) once the value of $f$
is known. We show the typical values of this parameter in
Figure~\ref{fig:fval} for several choices of $Z$.
\begin{figure}
  \includegraphics[width=15cm]{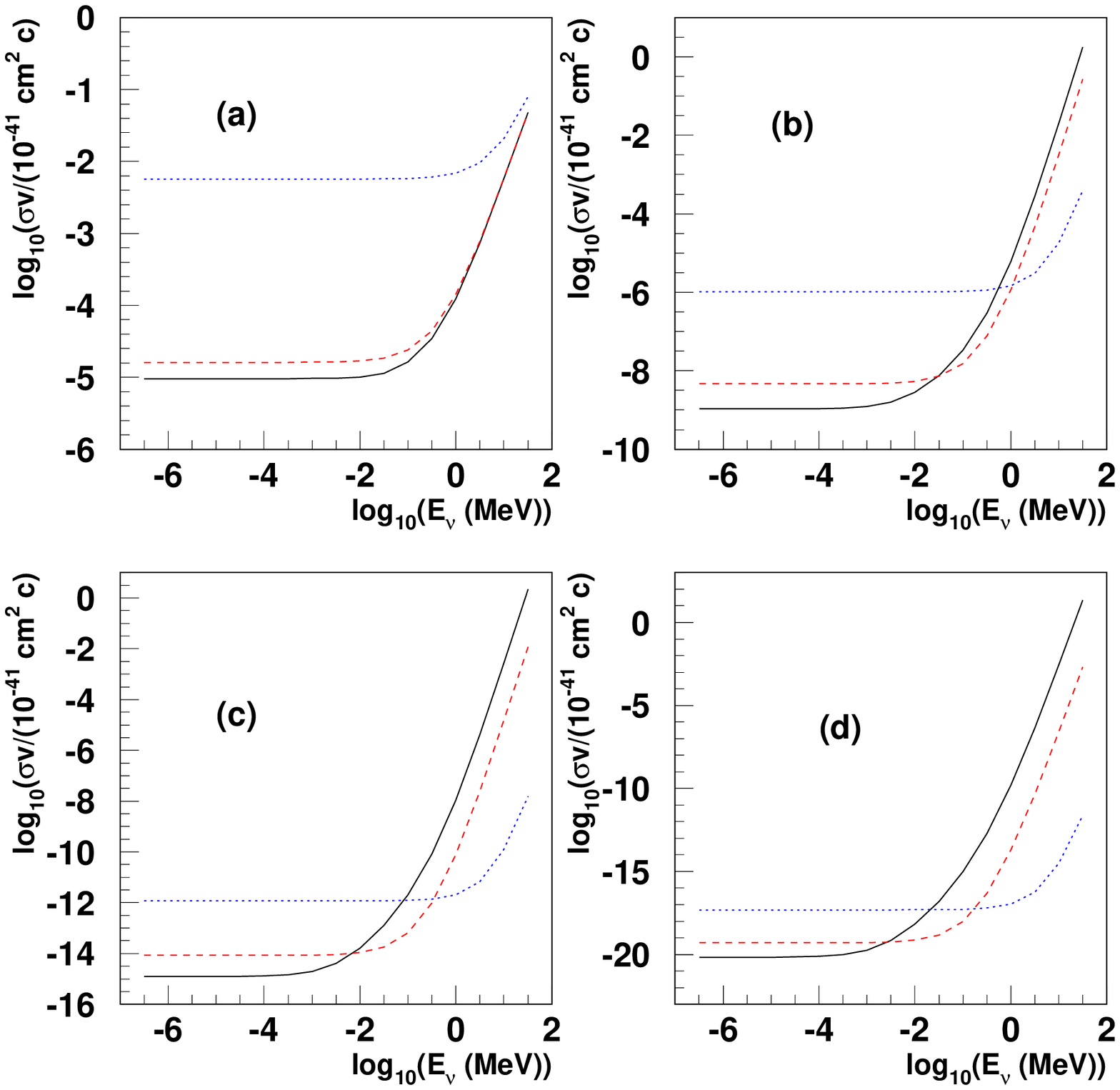}
\caption{\label{fig:xsect_m} The product
$\sigma_{\scriptscriptstyle {\rm NCB}} v_\nu$ for $\beta^-$
decaying nuclei versus neutrino energy. Typical values for
$\log(ft)$ values have been assumed~\cite{singh} as follows: (a)
allowed, $\log(ft_{1/2})=5.5$, (b) first unique forbidden,
$\log(ft_{1/2})=9.5$, (c) second unique forbidden,
$\log(ft_{1/2})=15.6$, (d) third unique forbidden,
$\log(ft_{1/2})=21.1$. The three curves refer to different
$Q_\beta$-values, solid line for $Q_\beta=10^{-3}$ MeV, dashed
line for $Q_\beta=10^{-1}$ MeV, dotted line  for $Q_\beta=10$ MeV.
Curves are for $Z=21$ and nuclear radius given by $R=1.2 A^{1/3}$
fm, where $A=2.5 Z$}
\end{figure}

\begin{figure}
  \includegraphics[width=15cm]{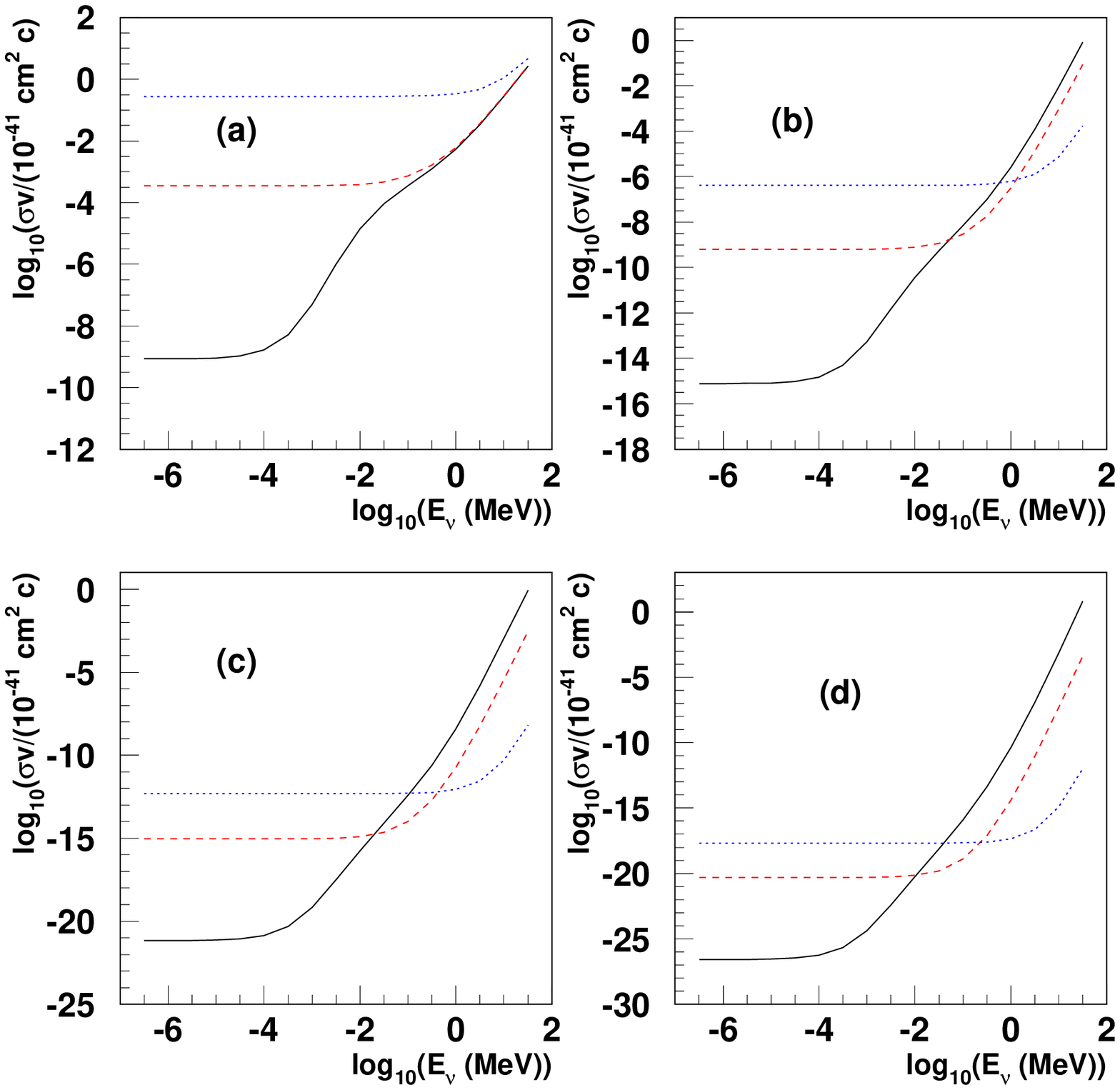}
\caption{\label{fig:xsect_p} The behavior of
$\sigma_{\scriptscriptstyle {\rm NCB}} v_\nu$ for $\beta^+$
decaying nuclei versus neutrino energy. We assume the following
values of $\log(ft_{1/2})$~\cite{singh}: (a) superallowed,
$\log(ft_{1/2})=3.44$, (b) first unique forbidden,
$\log(ft_{1/2})=9.5$ (c) second unique forbidden,
$\log(ft_{1/2})=15.6$, (d) third unique forbidden,
$\log(ft_{1/2})=21.1$. The three curves refer to different
$Q_\beta$-values, solid line for $Q=10^{-3}$ MeV, dashed line for
$Q_\beta=10^{-1}$ MeV, dotted line  for $Q_\beta=10$ MeV. Curves
are for $Z=21$ and nuclear radius given by $R=1.2 A^{1/3}$ fm,
where $A=2.5 Z$}
\end{figure}

\begin{figure}
  \includegraphics[width=15cm]{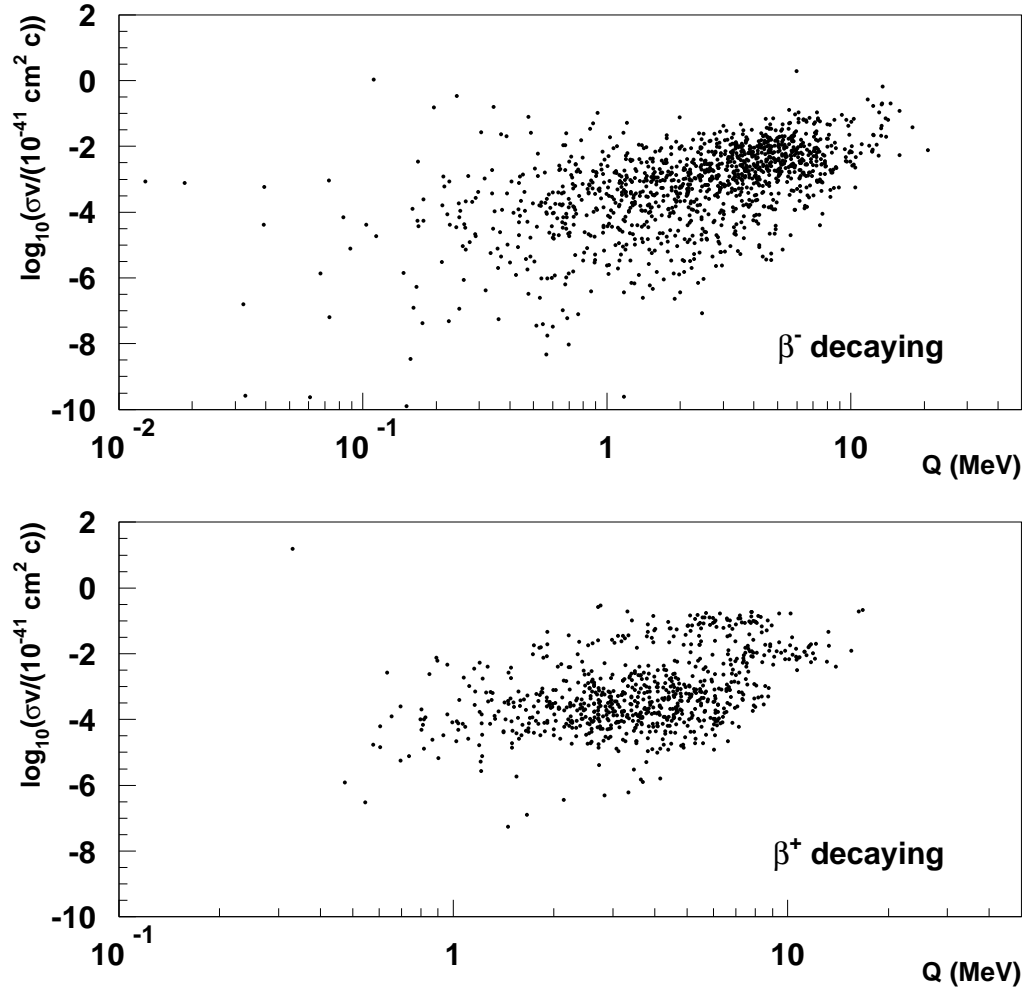}
\caption{\label{fig:xs_data} $\sigma_{\scriptscriptstyle {\rm
NCB}} (v_\nu/c)$ for $\beta^\pm$ nuclei as a function of the decay
$Q_\beta$-value and evaluated in the limit of low $p_\nu$. A total
of 1272 $\beta^-$ and 799 $\beta^+$ decays having strong $\beta$
branching ratio ($BR > 5\%$) is reported for both allowed and
forbidden decays}
\end{figure}

The behavior of $\sigma_{\scriptscriptstyle {\rm NCB}} v_\nu$
versus the incoming neutrino energy is shown in
Figures~\ref{fig:xsect_m} and \ref{fig:xsect_p} for $\beta^-$ and
$\beta^+$ decaying nuclei, respectively and for a particular
choice of $Z$. Notice that it reaches a plateau at low neutrino
energies, whose value strongly depends upon the kind of transition
(super-allowed, allowed, etc.) and $Q_\beta$. We have performed an
extensive calculation of the NCB cross section for all beta
decaying transitions listed in the ENSDF database~\cite{ensdf}. A
total of 14543 decays have been analyzed, 6288 $\beta^-$ and 8255
electron capture and $\beta^+$. We restricted our attention to
both allowed and unique forbidden decays having branching ratios
greater than $5\%$, namely 1272 $\beta^-$ decays and 799 $\beta^+$
decays. The corresponding $\sigma_{\scriptscriptstyle {\rm NCB}}
v_\nu$ for low incoming neutrino momentum are shown in
Figure~\ref{fig:xs_data}.
There are several nuclei spanning a wide range in $Q_\beta$ for
which interesting high values are reached in the range
$10^{-41}-10^{-43}$ cm$^2$. Some of these are reported in
Table~\ref{superp} were we collect the results for the best know
$0^+ \rightarrow 0^+$ superallowed decays. As explained in section
II, our estimate is based on the knowledge of experimental values
of $Q_\beta$ and $t_{1/2}$ which in this case are both known with
a great accuracy as this class of decays is used to test the CVC
hypothesis within the Standard Model~\cite{hardy}.

As we will discuss in details in the next Section, any use of NCB
processes to study low energy neutrino fluxes is crucially related
to the issue of rejection of background events represented by the
corresponding beta decay process. For a given incident neutrino
flux, the ratio of the NCB to decay events is proportional to
$\sigma_{\scriptscriptstyle {\rm NCB}} (v_\nu/c) \cdot t_{1/2}$ ,
so that nuclei with the highest value for this combination might
give the best chances if used in a future low energy neutrino
detection experiment. We report these nuclei in
Table~\ref{table:beta}. Among these, we recognize two isotopes,
namely $^3$H and $^{187}$Re, widely used in the past and ongoing
experiments or representing a future possibility for calorimetric
neutrino mass experiment \cite{fiorini}.
\begin{table}
\caption{\label{superp} The product $\sigma_{\scriptscriptstyle
{\rm NCB}} (v_\nu/c)$ for the best known superallowed $0^+
\rightarrow 0^+$ transitions. Numerical values for $Q_\beta$ and
partial half-lifes are taken from~\cite{hardy}. The value of $f$
is calculated adopting the parametrization of the Fermi function
of~\cite{wilk}.}
\begin{tabular*}{\textwidth}{@{}l*{15}{@{\extracolsep{0pt plus
12pt}}l}} \br Isotope &  $Q_\beta$ (keV)   & Half-life (sec) &
$\sigma_{\scriptscriptstyle
{\rm NCB}} (v_\nu/c)$ ($10^{-41}$ cm$^2$) \\
\mr
$^{10}$C       &  885.87  & $1320.99$ & $5.36\times 10^{-3}$ \\
$^{14}$O       & 1891.8   & $71.152$  & $1.49\times 10^{-2}$ \\
$^{26\rm m}$Al & 3210.55  & $6.3502$  & $3.54\times 10^{-2}$ \\
$^{34}$Cl      & 4469.78  & $1.5280$  & $5.90\times 10^{-2}$ \\
$^{38\rm m}$K  & 5022.4   & $0.92512$ & $7.03\times 10^{-2}$ \\
$^{42}$Sc      & 5403.63  & $0.68143$ & $7.76\times 10^{-2}$ \\
$^{46}$V       & 6028.71  & $0.42299$ & $9.17\times 10^{-2}$ \\
$^{50}$Mn      & 6610.43  & $0.28371$ & $1.05\times 10^{-1}$ \\
$^{54}$Co      & 7220.6   & $0.19350$ & $1.20\times 10^{-1}$ \\
\br
\end{tabular*}
\end{table}

\begin{table}
\caption{\label{table:beta} Beta decaying nuclei that present the
largest product of $\sigma{\scriptscriptstyle {\rm NCB}} (v_\nu/c)
\cdot t_{1/2}$ for low neutrino momentum and have a $\beta^\pm$
decay branching fraction larger than 80\%.}
\begin{tabular*}{\textwidth}{@{}l*{15}{@{\extracolsep{0pt plus
12pt}}l}} \br Isotope & Decay & $Q_\beta$  (keV) & Half-life (sec)
&
$\sigma_{\scriptscriptstyle {\rm NCB}} (v_\nu/c)$ ($10^{-41}$ cm$^2$) \\
\mr
$^{3}$H    & $\beta^-$ &  18.591 & $3.8878\times 10^8$    & $7.84\times 10^{-4}$ \\
$^{63}$Ni  & $\beta^-$ &  66.945 & $3.1588\times 10^9$    & $1.38\times 10^{-6}$ \\
$^{93}$Zr  & $\beta^-$ &  60.63  & $4.952\times 10^{13}$  & $2.39\times 10^{-10}$ \\
$^{106}$Ru & $\beta^-$ &  39.4   & $3.2278\times 10^7$    & $5.88\times 10^{-4}$ \\
$^{107}$Pd & $\beta^-$ &  33     & $2.0512\times 10^{14}$ & $2.58\times 10^{-10}$ \\
$^{187}$Re & $\beta^-$ &  2.64   & $1.3727\times 10^{18}$ & $4.32\times 10^{-11}$ \\
\hline & & & & \\
$^{11}$C  & $\beta^+$ &  960.2    & $1.226\times 10^{3}$ & $4.66\times 10^{-3}$ \\
$^{13}$N  & $\beta^+$ &  1198.5   & $5.99\times 10^{2}$  & $5.3\times 10^{-3}$ \\
$^{15}$O  & $\beta^+$ &  1732     & $1.224\times 10^{2}$ & $9.75\times 10^{-3}$ \\
$^{18}$F  & $\beta^+$ &   633.5   & $6.809\times 10^{3}$ & $2.63\times 10^{-3}$ \\
$^{22}$Na & $\beta^+$ &   545.6   & $9.07\times 10^{7}$  & $3.04\times 10^{-7}$ \\
$^{45}$Ti & $\beta^+$ &  1040.4   & $1.307\times 10^{4}$ & $3.87\times 10^{-4}$ \\
\br
\end{tabular*}
\end{table}

\section{NCB versus $\beta$ decay: the case of cosmological relic neutrinos}

We now consider a possible application of NCB to detection of the
background of cosmological relic neutrinos. Actually, this
represents one of the most ambitious challenges in modern
cosmology. There are two intertwined issues which should be
discussed, i.e. the event to background rate and the energy
resolution. First of all, though as we stressed several times the
NCB process is with no energy threshold, nevertheless, the ratio
of NCB event number to corresponding beta decay events is
typically very small. Using (\ref{sigma1}) and (\ref{sigma3}) we
find
\begin{equation}
\frac{\lambda_\nu}{\lambda_\beta} = \left(\lim_{p_\nu \rightarrow
0} \sigma_{\scriptscriptstyle {\rm NCB}} v_\nu \right) n_\nu
\frac{ t_{1/2}}{\ln 2}= \left(\lim_{p_\nu \rightarrow
0}\frac{2\pi^2}{{\cal A}} \right) \, n_\nu \, , \label{ratiotot}
\end{equation}
where we have used the fact that relic neutrinos have a very small
mean momentum of order $T_\nu$ with a spread of the same order of
magnitude, and the fact that the product of NCB cross section
times neutrino velocity gets an asymptotic constant value for
small neutrino energies, see Figures \ref{fig:xsect_m} and
\ref{fig:xsect_p}. In the case of $^3$H we get
\begin{equation}
\lambda_\nu(^3 {\rm H}) = 0.66 \cdot 10^{-23}\lambda_\beta(^3 {\rm
H}) \, .
\end{equation}
A numerical estimate of the relative event rate of low energy
neutrino NCB interaction with respect to beta decays can be
obtained by using the results of Figure~\ref{fig:aval} in case of
allowed and unique forbidden transitions.

Despite of this disappointing result, at least in principle the
experimental signature of NCB events is unambiguous as the
electron (positron) in the final state has a kinetic energy at
least $2m_\nu$ above the beta decay endpoint energy. However, the
finite energy resolution of any experimental apparatus and the
extremely low cross section make relic neutrino detection via NCB
a real challenge due to the large background events. In
particular, for low neutrino masses smaller than the typical
experimental energy resolution, it is really impossible to
disentangle the few expected NCB events from the large background
of standard beta events. In this case NCB processes are of no use.

On the other hand, in more optimistic scenario with comparable
values of neutrino masses and experimental energy resolution, the
situation could be much more promising. As an example, we consider
a future experiment reaching an energy resolution $\Delta$, and
neutrino masses in the eV range. From (\ref{sigma1}) and
(\ref{ratedecay}), the ratio of the event rate
$\lambda_\beta(\Delta)$ for the last beta decay electron energy
bin $W_o-\Delta<E_e<W_o$, compared with the total NCB event rate
can be easily calculated, giving
\begin{equation}
\frac{\lambda_\nu}{\lambda_\beta(\Delta)} = \frac{9}{2} \zeta(3)
\left( \frac{T_\nu}{\Delta} \right)^3 \frac{1}{\left(1+2
m_\nu/\Delta \right)^{3/2}}\, ,
\end{equation}
where we have used that $n_\nu = 3 \zeta(3) T_\nu^3/(4 \pi^2)$ and
the fact that $Q_\beta >> \Delta$. We have checked that this
expression is accurate at percent level for nuclei that undergo
both allowed and unique-forbidden transitions with endpoint energy
in the range $10^{-3} < Q_\beta < 10$ MeV. This gives for example,
the value $\lambda_\nu/\lambda_\beta(\Delta)\sim 2.2\cdot
10^{-10}$ for $\Delta = 0.2$ eV and $m_\nu = 0.5$ eV.

To get an estimate of the signal to background ratio we can
proceed as follows. For $\Delta < m_\nu$, the expected background
electron events which are produced by beta decay, yet having an
energy which corresponds to the relic neutrino capture energy bin
centered at $E_e=W_o+2 m_\nu$ is suppressed by the exponential
factor
\begin{equation}
\rho = \frac{1}{\sqrt{2 \pi}} \int_{2 m_\nu/\Delta-1/2}^{2
m_\nu/\Delta +1/2} \, e^{-x^2/2} dx \, .
\end{equation}
Thus, imposing a signal to background ratio larger than unity
corresponds to the condition
\begin{equation}
\frac{9}{2} \zeta(3) \left( \frac{T_\nu}{\Delta} \right)^3
\frac{1}{\left(1+2 m_\nu/\Delta \right)^{3/2} \rho}  \geq 1 \, ,
\end{equation}
and this for a given neutrino mass provides the energy resolution
which is necessary to achieve. A signal to noise ratio of order 3
is for example obtained if $\Delta= 0.2$ eV for $m_\nu = 0.7$ eV,
while a smaller neutrino mass of $0.3$ eV requires $\Delta=0.1$
eV. In these cases a total event number of order 10 is needed to
get a 5-$\sigma$ $discovery$ claim. Presently, this energy
resolution seems very hard to get. Nevertheless, if a large
neutrino mass will be found by ongoing beta decay experiments such
as KATRIN, it is not inconceivable that a future generation of
experiments might reach energy resolution as low as 0.1 eV.

Finally, we estimate the order of magnitude of the mass of
detector required to see neutrino events from the cosmological
background using NCB. For a mass $M[{\rm g}]$ expressed in grams,
the expected total event rate is
\begin{equation}
\lambda_\nu \frac{N_A\ M[{\rm g}]}{A} \, , \label{rate}
\end{equation}
where $N_A$ is the Avogadro number and $A$ is the atomic number of
the decaying nucleus. Inserting numerical values we get for the
molar rate
\begin{equation} 2.85 \cdot 10^{-2} \,
\frac{\sigma_{\scriptscriptstyle {\rm NCB}} v_\nu/c}{10^{-45} {\rm
cm}^2} \,{\rm yr}^{-1}\ {\rm mol}^{-1} \, . \label{exrate}
\end{equation}
As an interesting example, we consider the case of $^3$H. From
(\ref{exrate}) and using the results of Table \ref{table:beta}, we
estimate 7.5 events per year of data taking for a mass of 100 g.
On the other hand, a very small result is obtained in the case of
$^{187}$Re, due to the tiny NCB cross section,
$\sigma_{\scriptscriptstyle {\rm NCB}}(^{187}$Re$) v_\nu/c \sim
10^{-52}$ cm$^2$.

We conclude this Section by observing that all our findings are
obtained assuming a standard and homogeneous relic neutrino
background. However, massive neutrino density could be locally
larger because of gravitational clustering. This effect in a Cold
Dark Matter Halo could be relevant for order eV neutrino masses
\cite{singh2,wong}. Indeed,  the neutrino density in the
neighborhood of the earth is enhanced by a factor as large as
10$\div$20 for $m_\nu = 0.6$ eV or 3$\div$4 if $m_\nu=0.3$ eV
\cite{wong}. On the other hand, neutrino distribution in phase
space is very close to the standard homogeneous Fermi-Dirac result
for neutrino masses smaller than 0.1 eV. To quantify the effect of
gravitational clustering on NCB rate we again consider the
expected event rate for 100 g mass of $^3$H. Results are shown in
Table \ref{table:local}. The enhancement is simply due to the
larger integrated neutrino density $n_\nu$. Actually, even for
large neutrino masses the neutrino distribution has a mean
momentum of the order of 5$\div 8$ 10$^{-4}$ eV and decreases
exponentially for momenta larger than 10$^{-3}$ eV. This holds for
both a Navarro, Frenk and White and present day Milky Way mass
profiles, see Figure 6 in \cite{wong}. Therefore, the event rate
can be always computed as in (\ref{ratiotot}) in the limit of
vanishing neutrino momentum and taking into account the local
higher neutrino density $n_\nu$. This results in a remarkable
increase of the expected event number for large neutrino mass and
thus, for a fixed ratio $\Delta/m_\nu$, in a higher statistical
significance due to the larger signal to (beta decay) background
ratio.
\begin{table}
\caption{\label{table:local} The number of NCB events per year for
100 g of $^3$H, taking into account the effect of gravitational
clustering in the neighborhood of the earth, compared to the case
of a standard homogenous Fermi-Dirac distribution with $T_\nu =1.7
\cdot 10^{-4}$ eV (FD). We show for some value of neutrino mass
the results for a Navarro, Frenk and White profile (NFW) and for
present day mass distribution of the Milky Way (MW), using the
local neutrino densities computed in \cite{wong}.}
\begin{tabular*}{\textwidth}{@{}l*{15}{@{\extracolsep{0pt plus
12pt}}l}} \br $m_\nu$ (eV)  & FD (events yr$^{-1}$)&
NFW (events yr$^{-1}$)& MW (events yrs$^{-1}$)\\
\mr 0.6 & 7.5 & 90 & 150 \\
0.3 & 7.5 & 23 & 33 \\
0.15 & 7.5 & 10 & 12 \\
\br
\end{tabular*}
\end{table}
\section{Conclusions}

The detection of low energy neutrino backgrounds, primarily those
emitted at the freeze out of weak interactions in the early stages
of the evolution of the Universe, is still beyond our present
experimental capability. Nevertheless, there is a continuous
struggle both in the theoretical and experimental physics
communities in searching for new ways and methods to detect these
fleeting fluxes.

In this paper we have reported a careful analysis of neutrino
capture on beta decaying nuclei. These processes have the
remarkable property of having no energy threshold on the incoming
neutrino energy and thus they might represent a good and numerous
class of interactions suitable for low energy neutrino detection.
This idea was already suggested several years ago by S. Weinberg
in \cite{weinberg}, though in the framework of a degenerate
massless relic neutrino background. Our study has been inspired by
this and prompted by the fact that we presently know that
neutrinos are massive particles and might have masses $m_\nu$ as
large as a fraction of eV. Indeed, there is a gap of 2$m_\nu$ in
the electron or positron energy spectrum separating the (few)
events induced by e.g. relic neutrinos, from the large background
of standard beta decays. At least in principle, this allows to
disentangle the two processes. Of course, any possibility to
translate this idea into a real experimental technique depends
upon two crucial issues, namely the expected order of magnitude of
NCB event rate as well as the required energy resolution of the
outgoing electron (positron).

The first issue has been considered in details in this paper. We
have studied the low momentum limit of the NCB cross sections for
more than two thousands beta decaying nuclei with a branching
factor larger than 5$\%$. Whenever possible, we have tried to
reduce the effect of the uncertainty on the nuclear matrix
elements by linking the NCB rate to the experimentally known
properties of the corresponding beta decay, the half-life and the
$Q_\beta$ value. Depending on the involved nuclear transition, the
cross section times neutrino velocity in the low neutrino momentum
regime spans several order of magnitude and interestingly, can be
as large as $10^{-42} \div 10^{-43}$ cm$^2$ $c$ for super-allowed
decays. This means that the event rate can be remarkably large.
For example, we found that with 100 g of $^3$H one expects order
10 events per year due to scattering of cosmological relic
neutrinos. Similar rates are also found for many super-allowed
transitions. Interestingly, the event rate is expected to be even
larger for neutrino masses of 0.3 $\div$ 0.7 eV due to the effect
of gravitational clustering in the neighborhood of the earth. In
this case the expected signal rate is in the range 20$\div$150 per
year per 100 g of $^3$H.

The second problem is the possibility to have a reasonable
rejection of the background due to standard beta decay
electrons/positrons. This depends on the incoming neutrino energy
range which is under study and the experimental energy resolution.
In the most demanding case of the extremely low momentum
cosmological relic neutrinos, it is necessary to reach a
sensitivity which is better than the value of neutrino mass. This
can be foreseen as a plausible perspective for future experiments
only if neutrinos have masses of order eV, thus in the so called
degenerate scheme for neutrino masses, which is still allowed by
all present data, though slightly disfavored by cosmological
observations. For example, for $m_\nu \sim 0.5$ eV and an energy
resolution of $0.1 \div 0.2$ eV a reasonable event to background
discrimination is possible. For smaller neutrino masses it seems
very hard if not impossible at all to use NCB processes for relic
neutrinos.

Of course, further improvements of the background rejection could
be achieved by a careful and more complete reconstruction of the
kinematics of the observed events, including possibly the daughter
nucleus recoil energy. As a further example, for polarized nuclei
one could also measure the beta-gamma angular correlation in case
the daughter nuclei de-excite via gamma emission. However, these
detailed issues are beyond the aim of the present paper and might
deserve further insights depending on forthcoming information
about the absolute neutrino mass scale.

\ack

G. Mangano is pleased to thank the Galileo Galilei Institute,
Florence, Italy, for the hospitality and the INFN for partial
support during the early stages of this work. A.G. Cocco
acknowledges useful discussions with N. Lo Iudice.

\end{document}